\documentclass{jfm}
\usepackage{graphicx}
\usepackage{natbib}
\ifCUPmtlplainloaded \else
  \checkfont{eurm10}
  \iffontfound
    \IfFileExists{upmath.sty}
      {\typeout{^^JFound AMS Euler Roman fonts on the system,
                   using the 'upmath' package.^^J}%
       \usepackage{upmath}}
      {\typeout{^^JFound AMS Euler Roman fonts on the system, but you
                   dont seem to have the}%
       \typeout{'upmath' package installed. JFM.cls can take advantage
                 of these fonts,^^Jif you use 'upmath' package.^^J}%
      }
  \else
  \fi
\fi
\ifCUPmtlplainloaded \else
  \checkfont{msam10}
  \iffontfound
    \IfFileExists{amssymb.sty}
      {\typeout{^^JFound AMS Symbol fonts on the system, using the
                'amssymb' package.^^J}%
       \usepackage{amssymb}%
       \let\le=\leqslant  \let\leq=\leqslant
         
      }{}
  \fi
\fi
\ifCUPmtlplainloaded \else
  \IfFileExists{amsbsy.sty}
    {\typeout{^^JFound the 'amsbsy' package on the system, using it.^^J}%
     \usepackage{amsbsy}}
    {\providecommand\boldsymbol[1]{\mbox{\boldmath $##1$}}}
\fi
\newsavebox{\astrutbox}
\sbox{\astrutbox}{\rule[-5pt]{0pt}{20pt}}

\title[On the Reynolds-Orr energy equation]{Stability of the Couette-Poiseuille flow by the Reynolds-Orr energy equation}
\author[F. Lam]
{F.\ns  L\ls A\ls M}
\pubyear{1996}
\volume{538}
\pagerange{119--126}
\date{?? and in revised form ??}
\begin{document}
\maketitle
\begin{abstract}
The normal-mode analysis of the Reynolds-Orr energy equation governing the stability of viscous motion for general three-dimensional disturbances has been 
revisited. The energy equation has been solved as an unconstrained minimization problem for the Couette-Poiseuille flow. When the flow is subjected to two-dimensional streamwise disturbances, 
an exact solution for the vorticity can be obtained in the case of plane Couette flow. In addition, it is found that there exists a minimum Reynolds number ($R_{min}=42.32$) among the Couette-Poiseuille profiles. This Reynolds number is lower than the value by Orr ($44.3$) for plane Couette flow. Similarly, for two-dimensional spanwise disturbances, the lowest Reynolds number ($R_{min}=19.88$) has been found. For fully three-dimensional disturbances, it is shown that the minimum Reynolds number is in general {\itshape smaller} than the corresponding two-dimensional counterpart for all the Couette-Poiseuille profiles except plane Couette flow. In particular, the minimum Reynolds number for plane Poiseuille flow is $47.2$ and it is associated with a three dimensional disturbance whose spanwise wavenumber is about three times of the streamwise wavenumber. The present results show that the Squire's theorem does not hold for the basic flows other than the linear profile.
\end{abstract}
\section{Introduction}
Consider a basic flow of an incompressible viscous fluid with velocity vector $\mathbf{U}$, and let $u'$ denote the velocity vector of the perturbation.
The combined flow field $\mathbf{U}+u'$ is still solenoidal and its motion must be governed the Navier-Stokes equations and subject to the same boundary conditions as $\mathbf{U}$. \cite{Reynolds} and \citet[][pp.~122--138]{Orr} showed that the rate of change of the kinetic energy ($E$) of the perturbation is governed by 
\begin{equation} \label{eq:ReyOrr}
	-\frac{\mathrm{d} E}{\mathrm{d} t}= \int_{V} [\: u'_i u'_j S_{ij} + R^{-1} ( {\partial u'_i}/ { \partial x_j})^2  \:] \;\mathrm{d} \mathbf{x}
\end{equation}
where $S_{ij}=(\partial U_i/\partial x_j + \partial U_j /\partial x_i )/2$ is the rate of the strain tensor of the basic flow. All quantities are taken to be dimensionless. $\mathbf{x}=(x_1,x_2,x_3)=(x,y,z)$ denotes the computation domain. This equation is known as the Reynolds-Orr energy equation. The stability problems governed by the energy equation are also referred to non-linear problems because the perturbation flow is somehow arbitrary; in particular, its size relative to the basic flow is not necessarily infinitesimal. The $x$-axis or the streamwise direction is the direction of the basic flow. We deal with a fluid filling between two solid walls situated at $\pm d_w$. The origin of the co-ordinates is situated at the mid-way of the walls. The basic velocity of the Couette-Poiseuille flow is uni-directional and its normalized velocity is given by
\begin{equation} \label{eq:velprof}
  U(z) = \left\{
    \begin{array}{ll}
      A(1-z^2)+z, & 0{\le} A {\le} 1/2, \\[2pt]
      \sqrt{A} \: [\sqrt{A}(1-z^2)+2\: \sqrt{(1-A)} z],  & 1/2 {<} A {\le} 1
    \end{array} \right.
\end{equation}
for $-1{\le} z {\le} 1$. The Reynolds number $R={\mathbf{U}}_{max} d_w/\nu$. We assume that $u'$ is spatially periodic in the $x$ or $y$ directions so that the integration domain $V$ may extend to infinity in both directions, and the integration can then be taken over exactly one wavelength. If the Reynolds number $R$ of a fluid motion satisfies the condition
\begin{equation}
	R\:{<}\:{R_{min}}=\min_{u'{\neq}0, u'_i|_{\partial{V}}=0, \partial u'_i /\partial x_i=0 } \Big\{ \:  \Big[ \int_{V} \big( {\partial u'_i}/ { \partial x_j}\big)^2 \mathrm{d} \mathbf{x} \Big] \; / \; \Big[ - \int_{V} u'_i u'_j S_{ij} \mathrm{d} \mathbf{x} \Big] \: \Big \},
\end{equation}
then the motion of the basic flow is said to be asymptotically stable in the mean. In the above expression, the denominator represents a transfer of energy between the basic flow and the perturbation flow. To determine $R_{min}$ is equivalent to a variational problem. The corresponding Euler equation of the variation takes the form of
\begin{equation} \label{eq:euler}
	R^{-1} \Delta u'_i - u'_j S_{ij} = \partial p'/\partial x_i,\;\;\; \partial u'_i /\partial x_i=0,\;\;\; u'_i=0\; \mbox{on}\; \partial{V}. 
\end{equation}
The last statement is the no-slip boundary condition. The pressure $p$ plays the role of the Lagrange multiplier associated with the solenoidal constraint. 
Let $\alpha$ and $\beta$ stand for the disturbance wavenumbers in the $x$ and $y$ directions respectively. Consider the normal-mode of the form
\begin{equation}
	\{u'_1,u'_2,u'_3,p'\}(x,y,z) = \{f,g,h,q\}(z) \: e^{ \mathrm{i} \alpha x + \mathrm{i} \beta y },
\end{equation}
then the Euler equation reduces to the system of the differential equations
\begin{subeqnarray} \label{eq:euler-normal-mode}
  \gdef\thesubequation{\theequation \textit{a,b,c}}
   \sigma L f &-& U' h = \mathrm{i} \alpha \: q, \;\;\; \sigma L g = \mathrm{i} \beta \: q, \;\;\; \sigma   L h - U' f = q', \\
  \gdef\thesubequation{\theequation \textit{d,e}}
   \mathrm{i} \alpha f & + &  \mathrm{i} \beta g + h'=0, \;\;\; f(\pm1)=g(\pm1)=h(\pm1)=0
\end{subeqnarray}
where the operator $L= (\mathrm{d}^2/\mathrm{d} z^2-\gamma ^2) $, $\gamma^2=\alpha^2+\beta^2$ and  $\sigma^{-1}=R/2$. The primes stand for the differentiation with respect to $z$. 
After eliminating the components $g$ and $q$, we obtain
\begin{subeqnarray} \label{eq:ReyOrr3d}
  \gdef\thesubequation{\theequation \textit{a,b}}
   \sigma L^2 h = -\gamma^2 U' f - \mathrm{i} \alpha \: (U'h' + U''h), \;\;\; \sigma \gamma^2 Lf = \beta^2  U' h + \mathrm{i} \alpha \sigma \: (Lh')
\end{subeqnarray}
and the boundary conditions $f(\pm1)=h(\pm1)=h'(\pm1)=0$.
Based on the idea of energy dissipation of \cite{Reynolds}, Orr studied (\ref{eq:euler}) in an attempt to establish a criterion so as to supplement the stability obtained by solving the Orr-Sommerfeld equation. He restricted his calculations to the streamwise disturbances. For plane Couette flow, \cite{Joseph} pointed out that the minimum Reynolds number is associated with a spanwise disturbance. Stability criteria by the energy method have been established in \cite{Sharpe}, \cite{MacCreadie}, \cite{Busse} and \cite{JosephCarmi}. It should be noted that these papers solved the eigenvalue problems which are linear in $\sigma$ (or in $\sigma^2$). More background on the energy method can be found in \cite{Serrin} and \cite{Chandrasekhar}. The book by \cite{DrazinReid} contains a whole section on the role of the energy method. In the past three decades, there have been virtually no published works on the non-linear stability using the approach of the Reynolds-Orr energy. Our recent study shows that there exist some gaps in the knowledge of the stability for the Couette-Poiseuille flow.

The structure of (\ref{eq:euler-normal-mode}) is essentially different from that of the linearized Navier-Stokes equations (\citet[][\S 1]{Lin} or \citet[][\S 25]{DrazinReid}). In the linear theory, analysis has been facilitated by Squire's theorem, see \cite{squire}. However, Squire transform does not apply to (\ref{eq:euler-normal-mode}) except the case of the linear basic profile. In general we must consider fully three dimensional disturbances and consequently we have to deal with some non-linear eigenvalue problems. Nevertheless, the mathematical form of (\ref{eq:ReyOrr3d}) suggests that it may be advantageous to convert the equations into the equivalent integral equations. The Nystrom procedures can then be used to efficiently solve the integral equations. Such formulations convert the stability problems into the linear or quadratic matrix eigenvalue problems. It turns out that use of the integral equations simplifies the analytic treatment of the stability problems. In the present paper, we give a summary of the work developed from these ideas.
\section{Solution of (\ref{eq:ReyOrr3d}) for $\beta=0$}
For the two dimensional disturbances in the direction of the basic flow, (\ref{eq:euler-normal-mode}) or (\ref{eq:ReyOrr3d}) is simplified as the differential equation (\cite{Orr})
\begin{equation} \label{eq:ReyOrrStream}
L^2{h}(z)= - \lambda \: ( U' h'+ U'' h/2 ), \;\;\;
	h(\pm1)=h'(\pm1)=0
\end{equation}
where $\lambda=\mathrm{i} \alpha R$. The adjoint equation to (\ref{eq:ReyOrrStream}) takes the form of
\begin{equation} \label{eq:ReyOrrStreamAdj}
L^2 h^{\dagger}(z)= \lambda \: (U' {h^{\dagger}}' + U''{h^{\dagger}}/2),\;\;\; h^{\dagger}(\pm1)= {h^{\dagger}}'(\pm1)=0.
\end{equation}
\subsection{An exact solution for plane Couette flow} \label{sec:exactpcf}
We have used two methods to tackle (\ref{eq:ReyOrrStream}). The first one is particularly appropriate for plane Couette flow. Because of the linearity in the basic velocity profile, the differential equation (\ref{eq:ReyOrrStream}) is first converted into a Volterra integral equation of the second kind. In particular, the kernel $H$ of the resulting integral equation is in the convolution form. Consequently, the inverse Laplace transform must possess an analytic solution of closed form. The four linear independent solutions of the operator $L^2$ are found to be
\begin{subeqnarray}
  \gdef\thesubequation{\theequation \textit{a,c}}
   F_1(z)&=&e^{\alpha z},\; F_3(z) = -\lambda \{F_1(z)(z+\alpha_0)- \alpha_0 F_2(z) \}/2, \\
  \gdef\thesubequation{\theequation \textit{b,d}}
   F_2(z)&=& e^{-\alpha z},\; F_4(z)=-\lambda \{\alpha_0 F_1(z)+F_2(z) (z-\alpha_0) \}/2 
\end{subeqnarray}
where $\alpha_0=1/(-2 \alpha)$. Introducing the vorticity $\eta=Lh$, the method of variation of parameters is evoked to reduce (\ref{eq:ReyOrrStream}) to the equivalent Volterra integral equation 
\begin{equation} \label{eq:ReyOrr-volt}
\eta(z)=\sum_{j=1}^{4} c_j F_j(z) + \int_0^z H(z,s)\eta(s) \mathrm{d} s
\end{equation}
where $c_j$'s are arbitrary constants. The kernel is given by
\begin{equation}
H(z,s)= - \alpha \: \alpha_0^2 \: (z-s) \left( \:  e^{\alpha(z-s)} - e^{-\alpha(z-s)} \: \right).
\end{equation}
The special form of the kernel suggests that we seek the solutions of the integral equation (\ref{eq:ReyOrr-volt}) by means of the Laplace method. Denote ${\cal F}_j$'s and ${\cal H}$ as the Laplace transforms of $F_j$ and $H$ respectively. 
The solution of the integral equation is given by
\begin{equation}
\eta(z)=\sum_{j=1}^{4} c_j \left( F_j(z)+\frac{1}{2\pi \mathrm{i}} \int_{\gamma- \mathrm{i} {\infty}}^{\gamma+ \mathrm{i} {\infty}} e^{\sigma z}
{\cal F}_j(\sigma) \frac{\lambda \; {\cal H}(\sigma) }{1-\lambda \; {\cal H}(\sigma)} \mathrm{d} {\sigma} \right)
\end{equation}
where the value of $\gamma$ in the Bromwich contour of integration in the complex $\sigma$-plane 
is chosen so that all the singularities of the integrand lie on the left of the line, being parallel to the imaginary axis. The singularities of the integrand are determined by the four roots (assume distinct) of the quartic equation 
\begin{subequations} \label{eq:quatics}
\begin{equation} 
	\sigma^4-2 \alpha^2 \sigma^2  + \alpha^4 - 4\alpha d \sigma =0
\end{equation}
where $d=-\lambda \alpha \alpha_0^2$. Let $\sigma_k, k=1,2,3,4$ be the roots and they are readily found to be
\begin{equation}
\sigma_{1,2}=(\sqrt {6}P{\pm}Q)/6,\;\;\;
\sigma_{3,4}=-(\sqrt {6}P{\pm}Q)/6
\end{equation}
where
\begin{equation}
P=\sqrt{2\alpha^2+S+4\alpha^4/S},\;
Q=\sqrt {-{{(-24\,{\alpha}^{2}+6\,S+24\,{\alpha}^{4}/S+36\,d \alpha\sqrt {6}/P)}}}
\end{equation}
\end{subequations}
and 
$S=\sqrt[3]{8\alpha^6+27 d^2 \alpha^2 + 3 \alpha d \sqrt{48 \alpha^6 + 81 d^2 \alpha^2}}.$
The inverse transform can be carried out by means of partial fraction decomposition and the result is
\begin{equation} \label{eq:pcf-exact-sol}
\eta(z)=\sum_{j=1}^{4} c_j \left( F_j(z)+ \lambda  \sum_{k=1}^4 b_k \: \int_{0}^{z} e^{\sigma_k (z-s)} F_j(s) \mathrm{d} s \right)
\end{equation}
where $b_k=16 \alpha^4 \sigma_k  \prod_{j=1, j{\neq}k}^{4} (\sigma_k-\sigma_j)^{-1}$. (When $R=4/(3\sqrt{3}) \alpha$, the quartic equation admits repeated roots, there will be obvious modifications in the form of the solutions and the values of $b_k$.) The expression (\ref{eq:pcf-exact-sol}) constitutes an exact solution to the vorticity. Once $\eta(z)$ is known and so are $h(z)$ and $h'(z)$. The secular determinant, $\Delta(\alpha,R)=0$, is obtained by applying the four boundary conditions. 
\subsection{General velocity profile}\label{sec:velprof}
The first two of the linear independent solutions $F_1$ and $F_2$ remain unchanged while 
\begin{equation}
F_{3,4}(z)=-\lambda \: \Big(  \int_0^z Y(z,s) e^{ {\pm} \alpha s} \{ \alpha U'(s) + U''(s)/2 \} \Big) \mathrm{d}s
\end{equation}
where $Y(z,s)=\alpha_0 (e^{\alpha z} e^{-\alpha s} - e^{-\alpha z} e^{\alpha s})$ and $F_3$ takes the upper sign. Then the corresponding Volterra equation takes an alternative form of
\begin{subeqnarray} \label{eq:ReyOrr-volt-gen}
  \gdef\thesubequation{\theequation \textit{a}}
   \eta(z) &=& \sum_{j=1}^{4} c_j F_j(z) + \int_0^z (H_1+H_2)(z,s)\eta(s)\mathrm{d}s, \\
  \gdef\thesubequation{\theequation \textit{b}}
   H_1(z,s)&=& -\int_0^z \!Y(z,t) \frac{\partial Y(t,s)}{\partial s} U'(t) \mathrm{d}t,\\
  \gdef\thesubequation{\theequation \textit{c}}
   H_2(z,s)&=& - \int_0^z \! Y(z,t)Y(t,s) \: {U''(t)}/{2} \mathrm{d}t.
\end{subeqnarray}
Clearly the solutions to the integral equation will be much more involved. In the case of plane Poiseuille flow, we have
\begin{subeqnarray}
  \gdef\thesubequation{\theequation \textit{a}}
  H_1(z,s) &=& 4 \{\alpha^2 (z-s) (z+s) - \alpha (z-s) + 1 \}\sinh[\alpha(z-s)]\alpha_0^2, \\ 
  \gdef\thesubequation{\theequation \textit{b}}
	H_2(z,s) &=& \{2 \alpha (z-s) \cosh[\alpha(z-s)] - 2 \sinh[\alpha(z-s)]\}\alpha_0^2.
\end{subeqnarray} 
Note that $H_1$ is no longer in the convolution form.
\subsection{Reduction to Fredholm integral equation}
For non-linear velocity profiles, we prefer to use our second method because the self-adjoint differential operator $L^2$ together with its homogeneous boundary conditions has a simple inverse. Let the four linearly independent solutions of $L^2$ be
\begin{subeqnarray}
  \gdef\thesubequation{\theequation \textit{a,c}}
   \phi_1(z) &=& e^{\alpha z},\; \phi_3(z) = \alpha_0 \left\{ \psi_1(z) \phi_1(z)- \psi_2(z) \phi_2(z) \right\}, \\
  \gdef\thesubequation{\theequation \textit{b,d}}
   \phi_2(z) &=& e^{-\alpha z}, \;\phi_4(z) = \alpha_0 \left\{ \psi_3(z) \phi_1(z)- \psi_1(z) \phi_2(z) \right\}, \\
  \gdef\thesubequation{\theequation \textit{e,f,g}}
\psi_1(z) & =& z+1,\;
\psi_2(z) =\alpha_0 ( e^{-2 \alpha} -  e^{2\alpha z}),\; \psi_3(z) =\alpha_0( e^{-2 \alpha z} - e^{2\alpha}).
\end{subeqnarray}
\returnthesubequation
The Wronskian of the four solutions equals $4 \alpha^2$. The Green's function associated with $L^2$ satisfying the homogeneous Dirichlet boundary conditions is found to be
\begin{equation} \label{eq:ReyOrrGreens}
  G(z,y) = \alpha_0 \left\{\!\!
    \begin{array}{ll}
      \{\phi_2(y) + \Lambda(\alpha)[\psi_3(1)\phi_3(y)-\psi_1(1)\phi_4(y)]\}\phi_3(z)\; - \\
   \hspace {1cm} \{\phi_1(y) + \Lambda(\alpha)[\psi_1(1)\phi_3(y)-\psi_2(1)\phi_4(y)]\}\phi_4(z), & z \le y, \\
   & \\
      \{\phi_2(z) + \Lambda(\alpha)[\psi_3(1)\phi_3(z)-\psi_1(1)\phi_4(z)]\}\phi_3(y)\; - \\
   \hspace {1cm} \{\phi_1(z) + \Lambda(\alpha)[\psi_1(1)\phi_3(z)-\psi_2(1)\phi_4(z)]\}\phi_4(y),         & y \le z,
    \end{array} \right. 
\end{equation}
where $\Lambda(\alpha)=\alpha_0 /\{ \psi_2(1) \psi_3(1) - \psi_1^2 (1) \}$. Hence equation (\ref{eq:ReyOrrStream}) is equivalent to a homogeneous Fredholm integral equation of the second kind, namely
\begin{subeqnarray} \label{eq:ReyOrrFredIEStream}
  \gdef\thesubequation{\theequation \textit{a}}
  h(z) & + & \lambda \int_{-1}^{1} \! K(z,y) h(y) \mathrm{d} y=0, \\ 
  \gdef\thesubequation{\theequation \textit{b}}
	K(z,y)  &=& U'(y){\partial G(z,y)}/{\partial y} + {U''(y)}G(z,y)/{2}.
\end{subeqnarray} 
Note that $K(z,y)$ is non-symmetric. Since $G$ is defined in terms of linear combinations of the entire functions, it follows that  $|{K(z,y)}| {<}\infty$. In addition, the derivative of $K$ with respect to $y$ has finite jumps at $y=z$ and therefore it must also be bounded. Accordingly, the Fredholm determinant of (\ref{eq:ReyOrrFredIEStream}) is an entire function of $\alpha$ and $R$. Its order is at most $2/3$ for finite $\alpha$ and $R$. There exists at least one eigenvalue $R$ for every wavenumber $\alpha$. In fact, the spectrum consists of denumerable discrete eigenvalues. 
\section{Stability by the linear theory}
Our method of solution can also be applied to the Orr-Sommerfeld equation
\begin{equation} \label{eq:OS}
\lambda^{-1}L^2\phi= (U-c) (\phi''-\alpha^2 \phi) - U'' \phi, \;\;\ \phi(\pm1)=\phi'(\pm1)=0.
\end{equation}
In the temporal stability problem, $c$ is the complex eigenvalue to be determined via the eigenvalue relation $\Delta(\alpha,R,c)=0$. The equivalent Fredholm integral equation is
\begin{subeqnarray} \label{eq:OSFredIE}
  \gdef\thesubequation{\theequation \textit{a}}
  \lambda^{-1} \phi(z) & - & \int_{-1}^{1} \! K_L(z,y) \phi(y) \mathrm{d} y = -c \int_{-1}^{1} \! K_R(z,y) \phi(y) \mathrm{d} y, \\ 
  \gdef\thesubequation{\theequation \textit{b}}
	K_L(z,y) & = &  U\:LG(z,y) + 2 U' \:G'(z,y),\;\;K_R(z,y) =  L G(z,y).
\end{subeqnarray} 
The primes denote the differentiations with respect to $y$. Briefly, the Orr-Sommerfeld equation can be reduced to a generalized matrix eigenvalue problem when the integrals are approximated by quadratures. The second derivative of the basic flow does not play any role in the formulation. Hence the present method has the advantage in dealing with the velocity distributions that only the first derivative is available. 
It may be useful to calculate the critical Reynolds number by the spatial theory ($c$ is taken to be a purely real quantity. The wave number $\alpha$ is taken to be complex and is the unknown eigen-mode). Both the temporal theory and the spatial theory define the identical marginal stability; the critical Reynolds number is same in either theory. In addition, the spatial theory is mathematically ill-posed because the velocity and the pressure are unbounded at infinity (\cite{DrazinReid}). The ill-posedness applies to most basic flows. The reason why one may calculate many spatial modes is that the corresponding temporal modes exist. The converse is however not true in general; some eigen-modes of the linear stability are well-defined only in terms of the temporal theory. 
\section{Solution of (\ref{eq:ReyOrr3d}) for $\alpha=0$}
For the spanwise disturbances, the differential equation (\ref{eq:ReyOrr3d}) consists in a sixth order system. The Fredholm integral equation is found by considering the inverses of the operators $L^2$ and $L$ in (\ref{eq:ReyOrr3d}). The result is
\begin{subeqnarray} \label{eq:ReyOrrFredIESpan}
  \gdef\thesubequation{\theequation \textit{a}}
  \sigma^2  h(z) &=& - \beta^2 \int_{-1}^{1} G_0(z,y) \: U'(y) \: h(y) \mathrm{d} y, \\ 
  \gdef\thesubequation{\theequation \textit{b}}
	G_0(z,y) &=&  \int_{-1}^{1} G(z,s) F(s,y) U'(s) \mathrm{d} s
\end{subeqnarray} 
and $F$ is the Green's function for the operator $L$ satisfying the boundary conditions.
\section{General solution of (\ref{eq:ReyOrr3d})}
The sixth order system of the ordinary differential equations is reduced to the equivalent Fredholm integral equation 
\begin{subeqnarray} \label{eq:ReyOrrFredIEFull}
  \gdef\thesubequation{\theequation \textit{a}}
   \sigma^2 h(z) = - \beta^2  \int_{-1}^{1} \!  G_0  U'(y) h(y) \mathrm{d} y &+& \sigma \mathrm{i} \alpha \int_{-1}^{1} \!  \Big[ G_1 - \gamma^2 G_2 + \frac{\partial G}{\partial y} U'(y) \Big] h(y) \mathrm{d} y, \\
  \gdef\thesubequation{\theequation \textit{b}}
   G_1(z,y)& =&  \int_{-1}^{1} \!\!\! \partial^3 F(z,s) /\partial s^3 \: G(s,y) \: U'(s) \mathrm{d} s,\\
  \gdef\thesubequation{\theequation \textit{c}}
   G_2(z,y)& =&  \int_{-1}^{1} \!\!\! \partial F(z,s) /\partial s \: G(s,y) \: U'(s) \mathrm{d} s.
\end{subeqnarray}
\section{Computations}
\begin{figure}
  \centerline{\includegraphics[height=7cm,width=13cm,keepaspectratio]{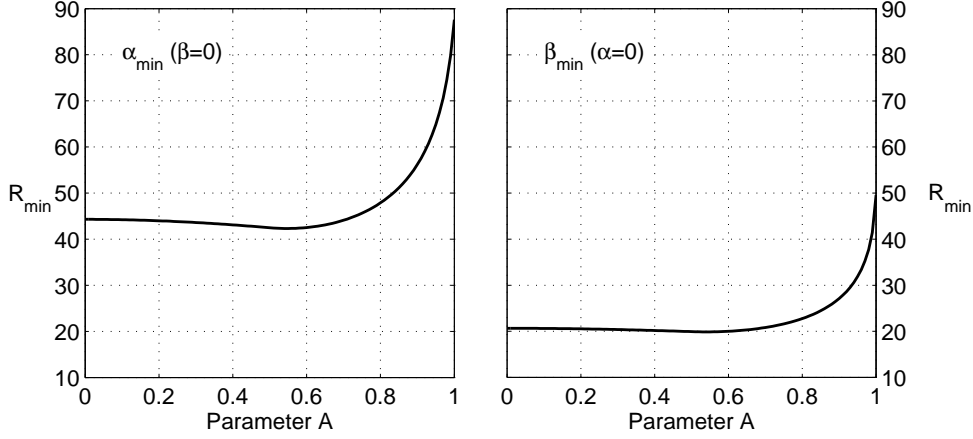}}
  \caption{Solutions of the Reynolds-Orr energy equation for the Couette-Poiseuille profile (\ref{eq:velprof}). $R_{min}$ at $\alpha_{min}$ and at $\beta_{min}$ is shown as a function of the basic flow parameter $A$. The case on the far left-hand side in each plot corresponds to plane Couette flow and at the far right-hand side plane Poiseuille flow.}\label{fig:RvsAB}
\end{figure}

\begin{figure}
  \centerline{\includegraphics[height=7cm,width=13cm,keepaspectratio]{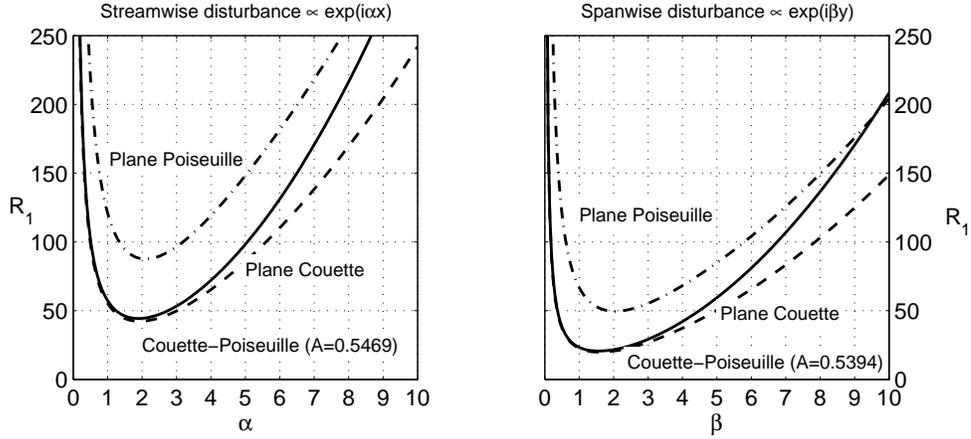}}
  \caption{Solutions of the Reynolds-Orr energy equations (\ref{eq:ReyOrrFredIEStream}) and (\ref{eq:ReyOrrFredIESpan}). The least Reynolds number $R_{1}$ of the spectrum at given wave number is shown as a function of ${\alpha}$ or ${\beta}$. The dash-dot lines are for plane Poiseuille flow, the solid lines for plane Couette flow and the dashed lines for the Couette-Poiseuille profile. By the minimization procedure, the lowest point in each curve defines the values of $R_{min}$ and $\alpha_{min}$ or $\beta_{min}$.}\label{fig:RvsWavenumber}
\end{figure}

\begin{figure}
  \centerline{\includegraphics[height=7cm,width=13cm,keepaspectratio]{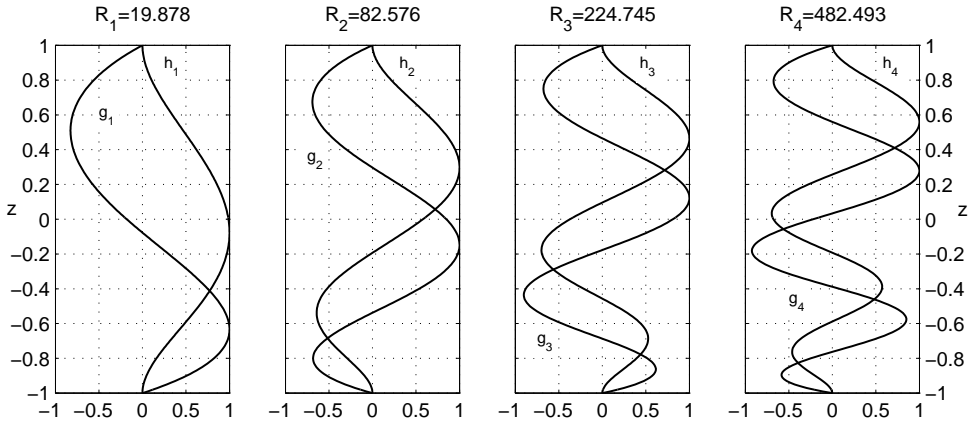}}
  \caption{Solutions of the Reynolds-Orr energy equation (\ref{eq:ReyOrrFredIESpan}) subject to spanwise disturbances $ \propto \exp({\mathrm{i} \beta y})$. From the left to right, the results of the first four modes of the eigen-functions at $\beta_{min}$ are plotted ($A=0.5394$).}\label{fig:EigFnSpanEx}
\end{figure}

Practically, the integral equations (\ref{eq:ReyOrrFredIEStream}) and (\ref{eq:ReyOrrFredIESpan}) have been solved by the standard Nystrom techniques. We have chosen the $N$-point Gauss-Legendre quadrature rule for the approximations since the kernels consist of well-behaved smooth functions. 
Some efforts have been given to the dependence of the accuracy of the eigenvalues on the Nystrom mesh. We denote the solutions of the integral equations by the epigraph optimization:
\begin{equation}
	\mbox{To minimize} \;\;\; R = R(\alpha,\beta, A)
\end{equation}
subject to no constraints. In particular, the solution of (\ref{eq:ReyOrrFredIEStream}) or (\ref{eq:ReyOrrFredIESpan}) reduces to the minimization problem of a univariate Reynolds-number function for fixed $A$ and to a bivariate function for variable $A$. To avoid excessive computations, we have made use of the theory of optimization which avoids any evaluations of the derivative of the Reynolds-number function.
For every given wavenumber $\alpha$ or $\beta$, the eigenvalue spectrum of $R$ is computed. The solutions of (\ref{eq:ReyOrrFredIEStream}) consist of complex conjugate pairs which give identical $u'_3$ while the spectrum of (\ref{eq:ReyOrrFredIESpan}) consists of simple eigenvalues. 
The variations of $R_{min}$ with the velocity parameter $A$ are plotted in Figure~\ref{fig:RvsAB}. Clearly there is a value of $A$ of the basic flow at which $R_{min}$ is least. The Reynolds-number functions are found to be convex and the epigraphs are shown in Figure~\ref{fig:RvsWavenumber}. The main numerical results are summarized in Table~\ref{tab:Stream}. In particular, plane Couette flow has been found stable for any $R\;{<}\;R_{min}\;=\;44.304\;\;\;\mbox{and}\;\;\;\alpha_{min}=1.8934$.
The symbol, $\alpha_{min}$, denotes the wavenumber at which the threshold Reynolds number occurs. The value of $R_{min}$ agrees with the result of Orr (1907) ($R=44.3$). Similarly, for purely spanwise disturbances, the numerical results are compared in Table~\ref{tab:Span}.  The eigen-functions of the spanwise disturbance at $A=0.5394$ are presented in Figure~\ref{fig:EigFnSpanEx}.
\begin{table}
  \begin{center}
\def~{\hphantom{0}}
  \begin{tabular}{lclll}
       Basic Flow (\ref{eq:velprof})  & $A$   & Author   &   $R_{min}$ & $\alpha_{min}$ \\
       & & & & \\
       Plane Couette: & 0 &\cite{Orr}  & 44.3 & 1.89 \\
       & 0 & Present  & 44.304 & 1.8934 \\
       Plane Poiseuille: & 1.0 & \cite{Orr}  & 87.75 & 2.1 \\
       & 1.0 & \cite{MacCreadie}  & 87.6 & 2.05 \\
       & 1.0 & Present & 87.594 & 2.0986 \\
       Couette-Poiseuille: & 0.5469 & Present & 42.320 & 1.9264 \\
  \end{tabular}
  \caption{Comparison of the results for streamwise disturbances $ \propto \exp({\mathrm{i} \alpha x})$}
  \label{tab:Stream}
  \end{center}
\end{table}
\begin{table}
  \begin{center}
\def~{\hphantom{0}}
  \begin{tabular}{lclll}
       Basic Flow (\ref{eq:velprof}) & $A$ & Author  &   $R_{min}$ & $\beta_{min}$ \\
       & & & & \\
       Plane Couette: & 0 & \cite{Pellew} & 20.663 & 1.5585 \\
       & 0 & Present     & 20.663 & 1.5582 \\
       Plane Poiseuille: & 1.0 & \cite{Busse}  & 49.604 & 2.044  \\
       & 1.0 & Present & 49.605 & 2.0437 \\
       Couette-Poiseuille: & 0.5394 & Present & 19.878 & 1.5936 \\
  \end{tabular}
  \caption{Comparison of the results for spanwise disturbances $ \propto \exp({\mathrm{i} \beta y})$}
  \label{tab:Span}
  \end{center}
\end{table}
It is known that the eigenvalue problem defined in (\ref{eq:ReyOrrFredIESpan}) is formally identical to the stability formulation of a viscous fluid contained between two rigid boundaries and heated from below, see \citet[][\S II]{Chandrasekhar}. The role of the Reynolds number plays that of the Rayleigh number. In addition, exact eigenvalue relations are available (see (46) and (51) of \cite{Pellew}). By the procedure of the unconstrained minimization, the critical Rayleigh numbers and the wavenumbers are then well defined and have been calculated. In the scalings of \cite{Pellew} or \cite{Reid}, we found the following:
\begin{subequations}
\begin{equation}
	\mbox{First even mode:}\;Ra_c=1707.7618,\;\beta_{min}=3.1163,
\end{equation}
\begin{equation}
	\mbox{First odd mode:}\;Ra_c=17610.3937,\;\beta_{min}=5.3646.   
\end{equation}
\end{subequations}
In comparison to the classical results, good agreement is obtained. The only noticeable difference occurs in the wavenumber of the even mode. (The value of $\beta_{min}=3.117$ has been quoted in most literature. Although it may appear to be a good numeral, it has been obtained by graphical method in the past in an arbitrary way because the ${R-\beta}$ graph was very ``flat'' in the vicinity of the minimum $\beta$.)

\begin{figure}
  \centerline{\includegraphics[height=6.75cm,width=13.5cm,keepaspectratio]{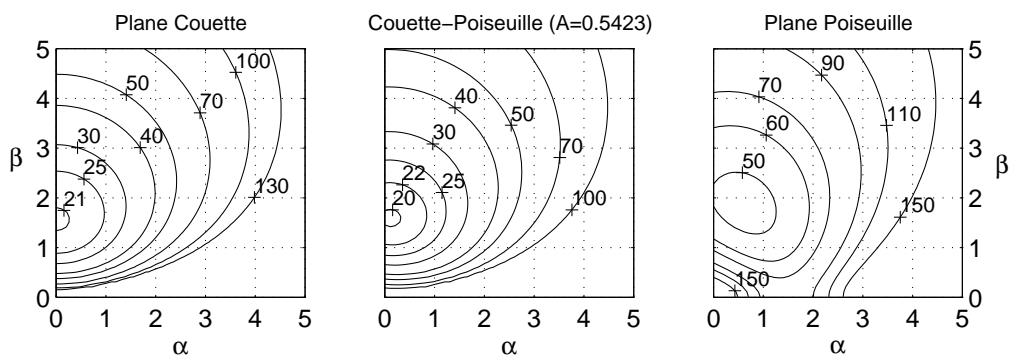}}
  \caption{The Reynolds number contours of the first mode from the solution of the Reynolds-Orr energy equation subject to fully three dimensional disturbances $\propto \exp({\mathrm{i} \alpha x + \mathrm{i} \beta y})$. In each illustration, $R_{min}$ is located at the centre of the Reynolds number basin.}\label{fig:R3DContours}
\end{figure}

\begin{figure}
  \centerline{\includegraphics[height=6.75cm,width=13.5cm,keepaspectratio]{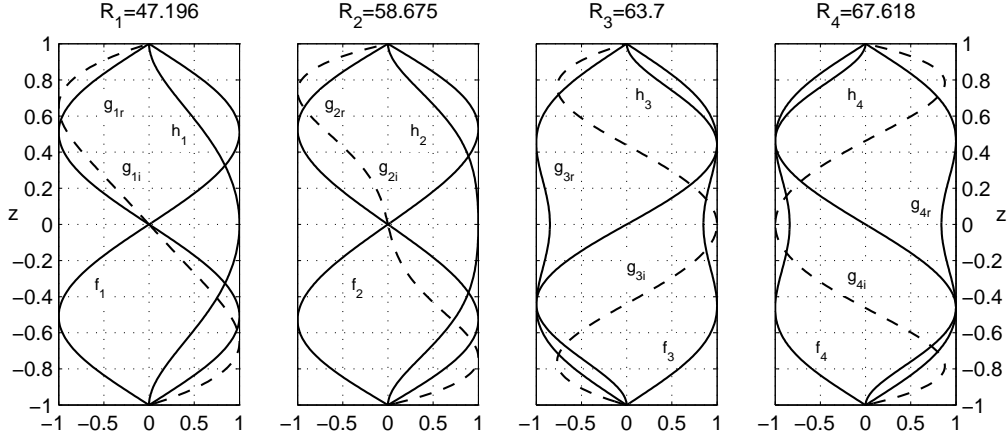}}
  \caption{The eigen-functions of the first $4$ modes at $\alpha_{min}$ and $\beta_{min}$ for plane Poiseuille flow subject to fully three dimensional disturbances $\propto \exp({\mathrm{i} \alpha x + \mathrm{i} \beta y})$. The Reynolds numbers of the modes are listed. Subscripts $r$ and $i$ are used to label the real and the imaginary parts respectively.}\label{fig:EigFnFull} 
\end{figure}

\begin{figure}
  \centerline{\includegraphics[height=7cm,width=13cm,keepaspectratio]{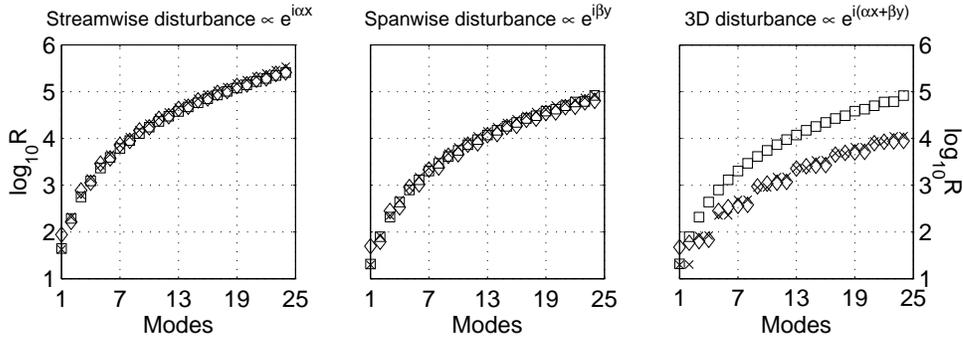}}
  \caption{The modal structure of the Reynolds-Orr energy equations (\ref{eq:ReyOrrFredIEStream}), (\ref{eq:ReyOrrFredIESpan}) and (\ref{eq:ReyOrrFredIEFull}) for the first 24 modes. In the left-hand side plot, the modal distribution is presented for $\alpha_{min}$. The symbols $\square$ are the modes for plane Couette flow, $\times$ for the Couette-Poiseuille profile of $A=0.5469$ and $\diamond$ for plane Poiseuille flow. The similar result is shown in the middle plot for $\beta_{min}$, $\times$ for $A=0.5394$. The results for fully three dimensional disturbances are shown in the right plot. The Couette-Poiseuille profile is for $A=0.5423$.}\label{fig:ModalDistEx}
\end{figure}

Application of the Nystrom method to (\ref{eq:ReyOrrFredIEFull}) leads to a quadratic eigenvalue problem of the form
\begin{equation} \label{eq:ReyOrrnonlinear}
(	\sigma^2 \mathbf{I} + \sigma \mathbf{A} + \mathbf{B}) \mathbf{h} = \mathbf{0} 
\end{equation}
where $\mathbf{A}$ and $\mathbf{B}$ are the discretization matrices, and $\mathbf{I}$ the identity matrix and $\mathbf{h}$ the eigen-function vector. The non-linear problem can be solved by means of the linearization technique of \cite{Wilkinson}. However, it is found that the non-linear problem is not completely reducible to the linear problem (\ref{eq:ReyOrrFredIEStream}) as $\beta {\rightarrow}0$ and $\alpha$ being large. Roughly speaking, no real Reynolds numbers of (\ref{eq:ReyOrrnonlinear}) are found for certain values of small $\beta$ and large $\alpha$. Since the results of the calculation for these wavenumbers do not compromise the main conclusion, no substantial efforts have been directed to resolve this issue. A selection of the computed results is presented in Figure~\ref{fig:R3DContours}. In the plot for Poiseuille flow, the minimum Reynolds number is clearly shown to be away from the axis of $\alpha=0$, indicating the three dimensional nature of the disturbance. The minimization problem for the other values of $A{>}0$ has been solved and a similar plot to the right-hand plot of Figure~\ref{fig:RvsAB} has been found. The main point of these computations is that every computed value of $R_{min}$ is lower than the corresponding value in the right-hand plot of Figure~\ref{fig:RvsAB}. Some numerical values of the minimum Reynolds number are listed in Table~\ref{tab:3d}.
\begin{table}
  \begin{center}
\def~{\hphantom{0}}
  \begin{tabular}{cccc}
       $Parameter A$ & $R_{min}$& $\alpha_{min}$ & $\beta_{min}$ \\
        & & & \\
       0.2 & 20.530 &    0.0457 &    1.5627  \\   
       0.4 & 20.150 &    0.0893 &    1.5756  \\   
       0.6 & 19.935 &    0.1315 &    1.5960 \\    
       0.8 & 22.603 &    0.1957 &    1.6417 \\ 
       1.0 & 47.196 &    0.6284 &    1.8381 \\
       0.5423 & 19.821 & 0.1185 & 1.5889 \\
  \end{tabular}
  \caption{Results of $R_{min}$ for fully three-dimensional disturbances $ \propto \exp({\mathrm{i} \alpha x} + {\mathrm{i} \beta y})$}
  \label{tab:3d}
  \end{center}
\end{table}
The eigen-functions for plane Poiseuille flow at $R_{min}$ are plotted in Figure~\ref{fig:EigFnFull}.
The modal distributions of the purely two-dimensional and the fully three dimensional disturbances are displayed in Figure~\ref{fig:ModalDistEx}. In every computation, we have examined the first few modes of the spectrum because the knowledge of these higher modes enables us to understand the critical mode.  
\section{Conclusions}\label{sec:concl}
We have shown how the technique of integral equation can be used to solve the Reynolds-Orr energy equation. Observation has been made on the mathematical structure of the equation for the special case of $\beta=0$. It is found that the case of plane Couette flow can be solved exactly because the kernel of the integral equation is in the convolution form. For general velocity profiles, analytic solutions of the problem become harder to obtain. The minimum Reynolds numbers obtained by \cite{Orr} for plane Couette and plane Poiseuille flows have been confirmed and they are numerically correct. For the purely two-dimensional disturbances, there exists a particular Couette-Poiseuille profile which gives rise to the lowest $R_{min}$ for $0{\leq}A{\leq}1$. For the three-dimensional disturbances, the Euler variation problem becomes a non-linear eigenvalue problem for minimizing the Reynolds number. It is shown that the minimum Reynolds number is in general smaller than the corresponding two-dimensional case for all the Couette-Poiseuille profiles. The only exception is plane Couette flow where the minimum Reynolds number is still related to the spanwise two-dimensional disturbance in the $yz$-plane. For plane Poiseuille flow, the minimum Reynolds number of the Euler equation is associated with a fully three-dimensional disturbance; $R_{min}$ is approximately $47.2$ with the streamwise wavenumber being one third of the spanwise wavenumber. The computational results show that the Squire's theorem is not valid for the basic flows except the linear profile.

The present study indicates that certain claims made in the existing literature (for instance \cite{DrazinReid}) on the stability characteristics of some basic flows by the Reynolds-Orr energy equation  are incomplete.
 
The threshold Reynolds numbers of the energy criterion are found ``rather low''. However the criterion has been derived exactly from the Navier-Stokes equations of motion. The non-linearity of the equations has been fully captured. 
On the other hand, the Cauchy problem for the Navier-Stokes equations in $\mathbf{R}^2$ (and in all likelihood in $\mathbf{R}^3$) has been shown to be well-posed. The property of the global regularity has the implication that the structural anomalies such as bifurcations and singularities have been ruled out. The velocity and the pressure are everywhere unique and smooth during the course of the flow development. The solutions of the linear theory (see \S 2) can be quantitatively assessed by direct comparisons with the Navier-Stokes solutions. In fact, some consequences implied by the linear theory are simply incompatible with the Navier-Stokes global regularity. In this note, we are mainly interested in the non-linear stability of the energy equation and hence we shall not elaborate on our assessment on the linear theory. For the initial data of practical interest, it has been known that the linear theory is unable to describe the complete evolution of fluid motion because the linearization fails to encapsulate the non-linearity which is the essence of fluid dynamics.
\begin{acknowledgments}
\end{acknowledgments}
\bibliographystyle{jfm}
\bibliography{ReynoldsOrr}
\vspace{0.5cm}
\begin{flushleft}
\texttt{f.lam11@yahoo.com} \\
\end{flushleft}
\end{document}